\documentclass[5p]{elsarticle}
\journal{Computational Materials Science}

\usepackage[utf8]{inputenc}
\usepackage{color}
\usepackage{comment}

\usepackage[colorlinks=true,allcolors=blue,pdfauthor=author]{hyperref}
\biboptions{sort&compress}

\usepackage{amsbsy}
\usepackage{amsfonts}
\usepackage{amsmath}
\usepackage{amssymb}
\usepackage{amsthm}
\usepackage{latexsym}
\usepackage{mathtools}

\usepackage{graphicx}
\usepackage{epsfig}
\usepackage{epstopdf}
\usepackage{svg}
\usepackage[caption=false]{subfig}

\usepackage{booktabs}
\usepackage{tabularx}

\renewcommand{\vec}[1]{\boldsymbol{#1}}
\providecommand{\mat}[1]{\boldsymbol{#1}}

\providecommand{\abs}[1]{\ensuremath{\left\vert#1\right\vert}}

\newcommand{\set}[1]{\left\{#1\right\}}

\providecommand{\about}[0]{\raise.17ex\hbox{$\scriptstyle\sim$}}

\DeclareMathOperator*{\argmin}{argmin}

\usepackage{lineno,hyperref}
\modulolinenumbers[5]
\definecolor{darkred}{rgb}{0.8, 0.0, 0.0}
\definecolor{darkgreen}{rgb}{0.0, 0.8, 0.0}
\definecolor{darkblue}{rgb}{0.0, 0.0, 0.8}

\providecommand{\vanish}[1]{}

\graphicspath{ {./figures/} }

\begin{document}

\begin{frontmatter}

\title{A Metric on the Polycrystalline Microstructure State Space}

\author[ucd]{Dylan Miley}
\ead{dcmiley@ucdavis.edu}

\author[ucd,nw]{Ethan Suwandi}

\author[gw]{Benjamin Schweinhart}

\author[ucd]{Jeremy K. Mason\corref{cor}}
\ead{jkmason@ucdavis.edu}

\address[ucd]{Department of Materials Science and Engineering, University of California, Davis, Davis, CA 95616, USA}

\address[nw]{Department of Materials Science and Engineering, Northwestern University, Evanston, IL 60208, USA}

\address[gw]{Mathematical Sciences Department, George Mason University, Fairfax, VA 22030, USA}

\cortext[cor]{Corresponding author}

\begin{abstract}
Material microstructures are traditionally compared using sets of statistical measures that are incomplete, e.g., two visually distinct microstructures can have identical grain size distributions and phase fractions.
While this is not a severe concern for materials fabricated by traditional means, the microstructures produced by advanced manufacturing methods can depend sensitively and unpredictably on the processing conditions.
Moreover, the advent of computational materials design has increased the frequency of synthetic microstructure generation, and there is not yet a standard approach in the literature to validate the generated microstructures with experimental ones.
This work proposes an idealized distance on the space of single-phase polycrystalline microstructures such that two microstructures that are close with respect to the distance exhibit statistically similar grain geometries in all respects below a user-specified length scale.
Given a pair of micrographs, the distance is approximated by sampling windows from the micrographs, defining a distance between pairs of windows, and finding a window matching that minimizes the sum of pairwise window distances.
The approach is used to compare a variety of synthetic microstructures and to develop a procedure to query a proof-of-concept database suitable for general single-phase polycrystalline microstructures.
\end{abstract}


\end{frontmatter}

\section{Introduction}
\label{sec:introduction}

One of the fundamental principles of materials science is that material properties are determined by microstructure.
For our purposes, microstructure is defined as material structure on the length scale of the grains or any minority phases, i.e., generally from tens of nanometers to hundreds of micrometers.
Structures at these length scales are readily characterized by electron microscopy techniques that provide micrographs with correlated chemical and crystallographic information, and the realization of integrated computational materials engineering (ICME) requires the ability to represent this wealth of information in a flexible and general way.
More specifically, a microstructure representation that enables meaningful comparisons of general microstructures would allow the correlation of processing routes with microstructures and of microstructures with material properties, the integration of disparate material datasets, the identification of alternative processing routes to realized microstructures, and the design of guided procedures to systematically explore unrealized ones.
This motivates the concept of an idealized microstructure state space with the following properties:
\begin{enumerate}
\item any microstructure can be represented as a point in the state space,
\item the state space specifies enough microstructural information to usefully constrain the material properties, and 
\item increasing the statistical similarity of two mi\-cro\-struc\-tures reduces the distance between the corresponding points in the state space.
\end{enumerate}
This purpose of this paper is to evaluate whether the concept of a microstructure state space could be practically useful for the quantitative comparison and classification of microstructures, at least for single-phase polycrystalline materials.
Our approach involves viewing a mi\-cro\-struc\-ture as a probability distribution of \emph{windows}, or microstructural volume elements.
This allows a \emph{metric}, or distance function, to be defined on the proposed microstructure state space by comparing probability distributions of windows.
An algorithm is proposed to approximately evaluate this metric, and is shown to be able to distinguish several classes of synthetic microstructures.
Moreover, only a small handful of windows is shown to be sufficient to correctly identify a microstructure in a proof-of-concept microstructure database.
Having achieved this paper's purpose, future work will extend the relevance of these ideas to more materials by including additional types of microstructural information in the proposed metrics.

\subsection{Probability Distributions of Windows}
\label{subsec:RVE}

For the concept of a microstructure state space to be useful, there needs to be consensus about precisely what information is required to specify a microstructure well enough to make meaningful property predictions.
A previously-proposed way to relate microstructure to properties is the representative volume element \cite{hill1963elastic,drugan1996micromechanics}, or the smallest volume of material for which constitutive relations relevant to the macroscopic material are sufficiently accurate for the intended purpose.
While this has the advantage of defining an explicit length scale above which microstructural features do not need to be considered, the length scale can be excessively large in practice. 
The concept of a statistical volume element (SVE) \cite{ostoja2006material,ostoja2007microstructural} instead recognizes that materials can be represented via a probability distribution of microstructural volume elements at a smaller user-specified length scale.
Increasing the length scale of the SVE provides more detail and tighter property bounds, but at the cost of increasing the difficulty of characterizing the probability distribution. 
Along the same lines, a point in the microstructure state space will be defined as a probability distribution of microstructural volume elements or \emph{windows}, with the space of windows 
being dependent on a user-specified length scale.

With this construction, microstructure representation only requires the specification of microstructure information below a particular length scale, but does not suggest a specific set of microstructural features sufficient to usefully constrain the material properties.
Moreover, the relevance of a given set of descriptors varies substantially across material systems;
for polycrystalline materials the microstructure is often closely associated with elements of the grain structure such as the grain size, aspect ratio, and orientation distribution, whereas for composite materials the phase distribution is often the most important feature \cite{yeong1998reconstructing,yeong1998reconstructingII,groeber2007framework,yin2008statistical}. 
The consequence is that selecting specific microstructural features relevant to any particular material risks making the microstructure representation incompatible with those for other material systems or classes \cite{furrer2019industrial}.
This not only limits the design space of materials to those for which the relevant microstructural features are those of known materials, but frustrates the construction of a general materials database that would ensure microstructure data is findable, accessible, interoperable, and reusable.

\subsection{Microstructure Descriptors}
\label{subsec:state_space_desc}

While finding a set of descriptors that can adequately characterize all known and unknown microstructures initially appears to be a daunting task, this is equivalent to finding a set of descriptors that is sufficient to reconstruct a microstructure (up to the variability inherent to materials systems) since any quantity of interest could then be measured from the reconstruction\footnote{Practically, this could resemble the image quilting approach to image synthesis in the computer science literature \cite{efros2023image}.}.
It is already well-established that the low-order microstructural descriptors mentioned above do not contain sufficient information to allow for a reliable reconstruction \cite{2004sundararaghavan,furrer2019industrial}, and hence do not fully describe a microstructure.
Accordingly, any microstructure representation that exclusively uses low-order descriptors would be unable to distinguish some distinct microstructures and would discard information that could be necessary to construct reliable property predictions in the future.

This motivates the exploration of high-dimensional descriptors designed to retain more information, a subject that has been approached in a variety of ways in the literature.
Correlation functions \cite{jiao2007modeling} and spectral density functions \cite{yu2017characterization} are both spatially-sensitive distributions, with the former specialized to multi-phase materials and the latter suited to materials with multiple features at different length scales.
For example, spectral density functions have been used to describe both the orientation distribution function and the geometry of grains belonging to a particular texture component \cite{fullwood2010microstructure}.
Other groups have used image analysis and computer vision techniques to convert selected regions of a secondary electron micrograph to sets of high-dimensional features by identifying, e.g., corners, edges, and intensity variations \cite{decost2017exploring}.
All of these approaches share the issue of being designed to encode a specific type of microstructure data, making the integration of disparate data types difficult due to their construction.

Machine learning approaches, and particularly convolutional neural networks, are increasingly being used to construct microstructure descriptors and can identify distinguishing features that are difficult even for experts to detect by hand.
These have the advantages of being relatively less subject to human bias and, since they generally operate directly on the micrographs, are applicable to a wide variety of materials.
Roughly speaking, convolutional neural networks operate by collecting information about the neighborhood of each pixel, aggregating and refining this information at successively longer length scales, and then feeding the resulting data through a standard neural net architecture to arrive at an overall description of the micrograph.
This description is constrained to be meaningful by a training process where the convolutional neural network is provided with a set of labeled (supervised) or unlabeled (unsupervised) training samples \cite{bostanabad2018computational,yang2018deepadversarial}.
Such approaches could eventually identify a set of distinguishing microstructural features suitable for general materials, but the existing literature is restricted to specific material classes where there have been systematic efforts to generate a sufficient number of high-quality micrographs \cite{decost2019high,muller2024overview}.
Moreover, it is unclear that general microstructural features generated in this way would be relevant or meaningful for materials that are not included in the training set, meaning that the use of such features could limit the possibilities for future materials design.
That said, these challenges could be partially addressed by using alternative neural network architectures particularly designed to investigate the properties the latent space (e.g., variational auto-encoders \cite{kingma2022autoencodingvariationalbayes}), since microstructures of materials between established material classes could then be investigated by interpolating on the latent space.

The related subject of synthetic microstructure generation and reconstruction is partly motivated by the need to have explicit microstructures to be able to computationally evaluate material properties within the ICME framework (synthetic microstructures can be generated much faster and at lower expense than the corresponding experimental ones).
The accuracy of the generated microstructures depends on both the information available in the microstructural descriptors and on the specifics of the generation algorithm, with a variety of options available depending on the material system, the required accuracy, and the computational cost \cite{chen2002phase,chun2020deep,kim2016two}.
For example, a grain structure can be generated at low cost by a Laguerre-Voronoi tessellation using an irregular sphere packing \cite{falco2017generation}, or at higher cost by a grain growth simulation that accounts for anisotropic grain boundary properties \cite{staublin2022phase}.
One widely-used option is the synthetic generation module of DREAM.3D \cite{Groeber2014dream} which employs a statistics-based approach.
DREAM.3D generates a polycrystal by iteratively evolving a packing of irregular ellipsoids until the corresponding grain structure sufficiently matches the desired statistics, allowing for a fine degree of control over the resulting microstructure. 
While this procedure is not associated with any particular physical process, it does allow for the generation of microstructures of widely varying statistics as part of an ICME pipeline.
Ideally, synthetic microstructures generated by any of these approaches would be validated by measuring the extent to which they resemble their experimental counterparts.
The expectation is that the two microstructures should map to the same point in the microstructure state space within statistical error, though the state space needs to be explicitly defined for this to be realized in practice.

\subsection{Utility of a Distance on Microstuctures}
\label{subsec:need_metric}

The ability to map a microstructure to a point in the state space, and to quantify the similarity of microstructures by the distance separating the corresponding points, would have widespread implications for materials science.
For example, this would allow one to precisely quantify the variability and reproducibility of microstructures in additively-manufactured metal parts over the entire domain of viable processing conditions \cite{franco2017sensory,dowling2020review}.
A materials database constructed on such a state space could allow users to interpolate the properties of a proposed material by means of a distance-weighted average of the properties of known materials.
Perhaps the most significant consequence from the standpoint of ICME though would be that the domain of such a state space would necessarily extend between and beyond known material classes, dramatically expanding the search space for future materials.
That said, there is prior literature that expresses the belief that finding a set of microstructure descriptors suitable for general material reconstruction is an intractable problem, and that a situational approach is the only feasible option \cite{bostanabad2018computational}.
Our belief is that this is overly conservative, and that descriptors can be formulated that are suitable for a more diverse set of materials than has been done previously.

Our intention in this paper is to develop a set of descriptors that characterize the geometry of grain structures in general polycrystalline materials, initially disregarding orientation, composition, and phase information.
Two microstructures with grain structures that are indistinguishable in this regard should be statistically identical in every geometric respect below a user-specified length scale.
This is achieved by viewing a microstructure as a probability distribution of windows---effectively an SVE---and quantifying the similarity of the probability distributions of windows;
prior literature on SVEs is generally concerned with the prediction of material properties rather than with the quantitative comparison of the SVEs themselves \cite{park2022microstructure, pilgar2023microstructure}.
After detailing the construction of the state space further in Sec.\ \ref{sec:state_space} and a way to measure distance on the state space in Sec.\ \ref{sec:wasserstein}, a way of quantifying the similarity or dissimilarity of windows is proposed in Sec.\ \ref{sec:window_comparison}.
This is used with the machinery developed in the preceding sections to evaluate the data requirements of the proposed approach and to quantify the similarity of a variety of synthetic microstructures in Sec.\ \ref{sec:comparing_materials}, and to develop a viable procedure to query a database of general single-phase polycrystalline materials in Sec.\ \ref{sec:database}.

\section{Microstructure State Space}
\label{sec:state_space}

Given the absence of consensus in the literature about what should constitute a microstructure state space, it is useful to be as clear as possible at the outset about what is required.
Our belief is that the three properties indicated at the beginning of Sec.\ \ref{sec:introduction} are essential, and that any proposed state space should be evaluated by the extent to which these criteria are satisfied.

\subsection{Contextualization of State Space Properties}
\label{subsec:contextualization}

With regard to the first property, that any microstructure can be represented as a point in the state space, one subtlety is that many instantiations of what are nominally equivalent microstructures should be identified with the same point up to statistical error.
For example, consider a facility that manufactures steel plates using a fixed and standardized processing procedure.
It is fantastically unlikely that there exist two substantial volumes of material with precisely the same microstructures in any of the steel plates manufactured over the lifetime of the facility.
That said, standardized processing ensures that the microstructures of such volumes of material are sufficiently similar that the resulting material properties are highly reproducible.
This implies that all of the microstructures produced by a standardized processing procedure should be identified with the same point in the state space up to statistical error, and that a suitable notion of equivalence should reflect their statistical similarity.
Along the lines proposed in Sec.\ \ref{sec:introduction}, a standardized processing procedure is viewed as being associated with a probability distribution of microstructural volume elements in the steel plates at a user-specified length scale.
That is, a processing procedure is associated with an SVE (i.e., a probability distribution of windows) that can be used to predict the properties of the steel plates, with an individual steel plate containing volume elements that sample from the probability distribution.
This approach has the advantage that material standards defined using SVEs instead of processing procedures would allow manufacturers to use any processing procedure for which the resulting microstructure conforms to the specification, a flexibility that would be valuable in the age of advanced manufacturing.

\begin{figure}
    \centering
    \includegraphics[width=\columnwidth]{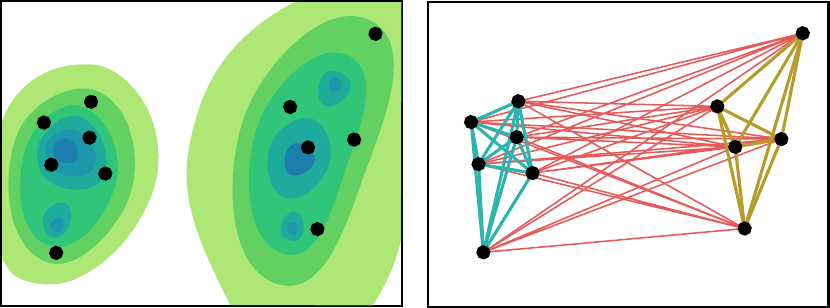}
    \caption{\label{fig:window_space}Two probability distributions on the window space for two distinct materials can be approximated by points clouds of sampled windows (left).
    Even without a coordinate system, the shape of a point cloud can be inferred from the pairwise distances between the points (right, blue and yellow lines).
    Moreover, the similarity of the underlying probability distributions for two materials can be inferred from the pairwise distances between points in the two clouds (right, red lines).}
\end{figure}

The second property of the state space, that enough microstructural information be specified to usefully constrain the material properties, requires that the representation of the SVE be explicit enough to allow for property predictions.
Since an SVE is fundamentally a probability distribution, it is necessary to know the space on which this probability distribution is defined;
this is precisely the space of all microstructural volume elements or the \emph{window space}.
Given the complexity of the window space, defining a coordinate system that would allow the probability distribution of windows (or a \emph{window distribution}) to be written down as an explicit function would be difficult.
Fortunately this is not necessary.
For many purposes a coordinate system can be replaced by a distance function (or a \emph{metric}) and a probability distribution can be replaced by a point cloud that sufficiently samples from the probability distribution.
A suitable point cloud can be constructed by sampling from micrographs generated by any one of a variety of characterization techniques, though not all characterization techniques provide data of the same type.
This possible heterogeneity of data means that the metric on the window space should be a flexibly-defined function that allows the user to specify precisely what features of the windows should be compared, but otherwise satisfies the usual (pseudo-\footnote{Two windows that are the same with respect to the selected data could be different with respect to the unselected data.})metric axioms:
\begin{enumerate}
\item the distance between a window and itself is zero,
\item the distance between two windows is symmetric, and
\item the triangle inequality is satisfied.
\end{enumerate}
Given such a function, the shape of a point cloud is completely specified by the set of all pairwise distances between points in the cloud; a graphical representation of this idea is given in Fig.\ \ref{fig:window_space}.
Moreover, a metric on the window space (e.g., the one developed in Sec.\ \ref{sec:window_comparison}) could in principle be used to define a coordinate system if necessary via manifold learning techniques.

The third property relates to the measurement of distances on the state space.
As is described below, a metric on windows can be used to construct a metric on microstructures.
Before discussing the properties of the microstructure metric, it is useful to briefly consider several example applications that are essential to the realization of ICME and would be enabled by such a function.
First, the ability to precisely quantify the similarity of two microstructures by a single number would allow rigorous verification that materials conform to microstructure-based standards, and that physics-based simulations of microstructure evolution accurately reproduce experimental observations.
Second, a metric could allow the construction of processing pathways to arbitrary target microstructures by, e.g., using a pattern-search minimization algorithm with the objective function being the distance to the target microstructure.
That is, one could consider an iterative process where the next step in the processing pathway is chosen by applying each of a variety of processing procedures to samples of the current material and selecting the one that most rapidly reduces the distance to the target.
Third, a metric would allow materials properties to be interpolated on the state space using standard techniques where the properties of an unrealized microstructure are predicted using the properties of nearby realized microstructures.
Such interpolations would be systematically improvable, with the possibility of identifying regions where the uncertainty is high, experimentally or computationally testing the properties of a material with that microstructure, and updating the interpolation function.

\subsection{A Metric for the State Space}
\label{subsec:metric_state_space}

Given that points on the state space define microstructures as window distributions, the desired metric that measures distance between points on the state space is equivalent to a function that measures the similarity of probability distributions on the window space.
Of the many metrics for probability distributions that have already been proposed in the literature, the Wasserstein metric \cite{kantorovich1960mathematical,villani2009optimal} stands out in several respects.
First, the metric is intuitively defined as the minimum cost to transform one probability distribution into the other, where the cost of transporting a probability mass is the magnitude of the probability mass times the transportation distance.
Second, the definition allows for the probability distribution to be continuous or discrete, a feature that is particularly useful since the continuous probability distribution of the SVE is practically replaced by a discrete population of sampled windows.
Specifically for the microstructure state space, computation of the Wasserstein metric effectively reduces to finding the best match between two populations of sampled windows, where the cost of matching a single pair of windows is equivalent to the distance between them on the window space.
As a result, the computation of a distance on the state space is reduced to the computation of pairwise distances on the window space and an optimization problem that has been extensively studied in the literature \cite{villani2009optimal}.

\begin{figure}[t]
    \centering
    \includegraphics[width=\columnwidth]{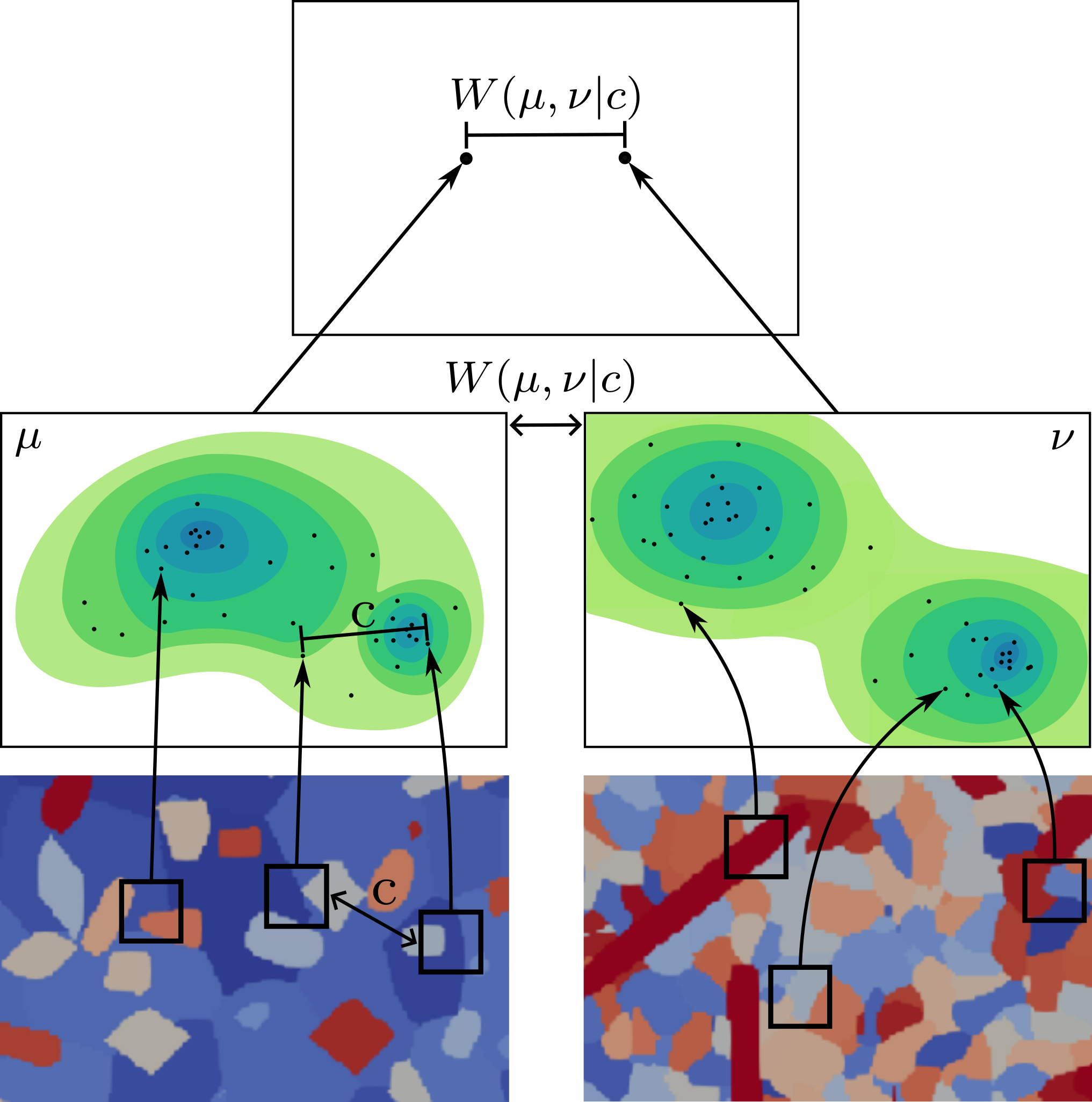}
    \caption{\label{fig:all_spaces}A schematic showing the various concepts involved in the construction of the microstructure state space.
    Windows of a consistent size are sampled uniformly at random from the interiors of the micrographs on the bottom row.
    The windows are represented as points in the window spaces on the middle row, with the points distributed according to underlying probability distributions $\mu$ and $\nu$ and distances on the window space given by $c$.
    Window distributions are represented as points in the microstructure state space on the top row, with the distances on the state space given by $W(\mu, \nu | c)$ (the notation is introduced in Sec.\ \ref{sec:wasserstein}).}
\end{figure}

Figure \ref{fig:all_spaces} gives a visual representation of these ideas.
Two micrographs of material microstructures appear on the bottom row, with windows of a consistent size sampled uniformly at random from the interiors of the micrographs.
The windows are represented as points in the window space on the middle row, with the points distributed according to underlying probability distributions $\mu$ and $\nu$ that characterize the two microstructures.
While there is only a single window space, the two probability distributions are shown on two copies of this space for visual clarity.
Distances between points in the window space are measured by a function $c$ that quantifies the similarity of the corresponding windows.
The window distributions are represented as two points in the microstructure state space on the top row.
Distances between points in the state space are measured by the Wasserstein metric $W(\mu, \nu | c)$ that quantifies the similarity of the corresponding window distributions for the window metric $c$.
That is, the statistical similarity of the two microstructures on the bottom row at the length scale of the windows is quantified by the distance between the corresponding points in the microstructure state space on the top row.

While the general approach of defining a probability distribution of local features and quantitatively comparing materials by means of these probability distributions has appeared in the literature before in the context of grain boundaries \cite{schweinhart2016topological} and silica glasses \cite{schweinhart2020statistical}, the definition of the windows and the metric on the window space for this paper is quite different.
As mentioned above, the definition of the metric function will depend on precisely which features of the windows are being compared.
For a proof of concept only the geometry of the grain structure is considered here, with the comparison of other microstructural features (e.g., composition or crystallographic orientation) relegated to the future work.
Since the proposed metric on the window space also makes use of the Wasserstein metric, a introduction is provided below.

\section{Wasserstein Metric}
\label{sec:wasserstein}


A fundamental problem in logistics is finding the most efficient way to transport goods from the points of production to the points of consumption.
Let $X = \{x_1, ..., x_{N}\}$ be a set of producers and $Y = \{y_1, ..., y_{M}\}$ be a set of consumers.
The producers have supplies $\vec{\mu} = (\mu_1, ..., \mu_{N})$ of goods and the consumers have demands $\vec{\nu} = (\nu_1, ..., \nu_{M})$ for goods, subject to the constraint that the total supply is equal to the total demand or $\sum_i \mu_i = \sum_j \nu_j$.
The incremental cost $c_{ij}$ of transporting a unit of goods from producer $x_i$ to consumer $y_j$ is assumed to be proportional to the intervening distance.
A \emph{transport plan} indicates the numbers of goods $\gamma_{ij}$ to be shipped from each producer $x_i$ to each consumer $y_j$, subject to the constraints
\begin{equation}
\gamma_{ij} \geq 0 \qquad \sum_{j} \gamma_{ij} = \mu_{i} \qquad \sum_{i} \gamma_{ij} = \nu_{j}.
\label{eq:coupling}
\end{equation}
That is, the number of goods shipped along any given route cannot be negative, the entire supply of each distributor is shipped, and the entire demand of each consumer is satisfied.
If $\Gamma$ is the set of all possible transport plans that satisfy these constraints, then the minimum cost transport plan $\mat{\gamma}^*$ is defined by the equation
\begin{equation}
\mat{\gamma}^* = \argmin_{\substack{\mat{\gamma} \in \Gamma}} \sum_{ij} c_{ij} \gamma_{ij}.
\label{eq:optimal}
\end{equation}
Finding $\mat{\gamma}^*$ is known as the optimal transport problem in the literature \cite{villani2009optimal}.

The optimal transport problem appears in a variety of contexts, and while the equations are usually the same the quantities involved can have different names.
Suppose that $X$ and $Y$ are two sets of points on an underlying space $Z$ equipped with a metric $d(x, y)$.
A \emph{discrete probability distribution} on $Z$ is a set of points with an associated set of nonnegative values (a measure) subject to the constraint that the sum of all the values is one.
Let $\vec{\mu}$ and $\vec{\nu}$ be discrete probability distributions associated with $X$ and $Y$.
A \emph{coupling} $\mat{\gamma}$ between $\vec{\mu}$ and $\vec{\nu}$ is a matrix with entries $\gamma_{ij}$ that satisfy the constraints given in Eq.~\ref{eq:coupling}.
Given a nonnegative cost matrix $\mat{c}$ with entries $c_{ij}$, the quantity $W(\vec{\mu}, \vec{\nu} | \mat{c}) = \sum_{ij} c_{ij} \gamma^*_{ij}$ is the cost of the optimal coupling in Eq.~\ref{eq:optimal}.
The function $W(\vec{\mu}, \vec{\nu} | \mat{c})$ then describes the similarity of $\vec{\mu}$ and $\vec{\nu}$ (the cost of transforming $\vec{\mu}$ into $\vec{\nu}$), and moreover has the properties of a metric when the entries $c_{ij}$ of the cost matrix are equal to the distances between the corresponding points $x_i$ and $y_j$ \cite{villani2009optimal}.
That is, given a metric $d(x, y)$ that measures distances between points in $Z$, the function $W(\vec{\mu}, \vec{\nu} | \mat{c})$ is a metric that measures the similarity of discrete probability distributions on $Z$.

The relevance of this to distances on the microstructure state space is established in the following way.
Let $Z$ be the window space and $X$ and $Y$ be sets of windows respectively sampled from a first and second micrograph.
The vectors $\vec{\mu}$ and $\vec{\nu}$ are defined to be uniform discrete probability distributions on $X$ and $Y$, i.e., $\mu_i = 1 / |X|$ and $\nu_i = 1 / |Y|$ where $|A|$ is the number of points in the set $A$.
Given a metric $c_{ij} = d(x_i, y_j)$ on the window space that measures the distance between windows $x_i$ and $y_j$, the quantity $W(\vec{\mu}, \vec{\nu} | \mat{c})$ then indicates the similarity of the sets of windows $X$ and $Y$.
That is, provided that the sampling of windows is sufficient to characterize the micrographs, $W(\vec{\mu}, \vec{\nu} | \mat{c})$ is the desired distance on the microstructure state space.

\begin{figure}
    \centering
    \includegraphics[width=\columnwidth]{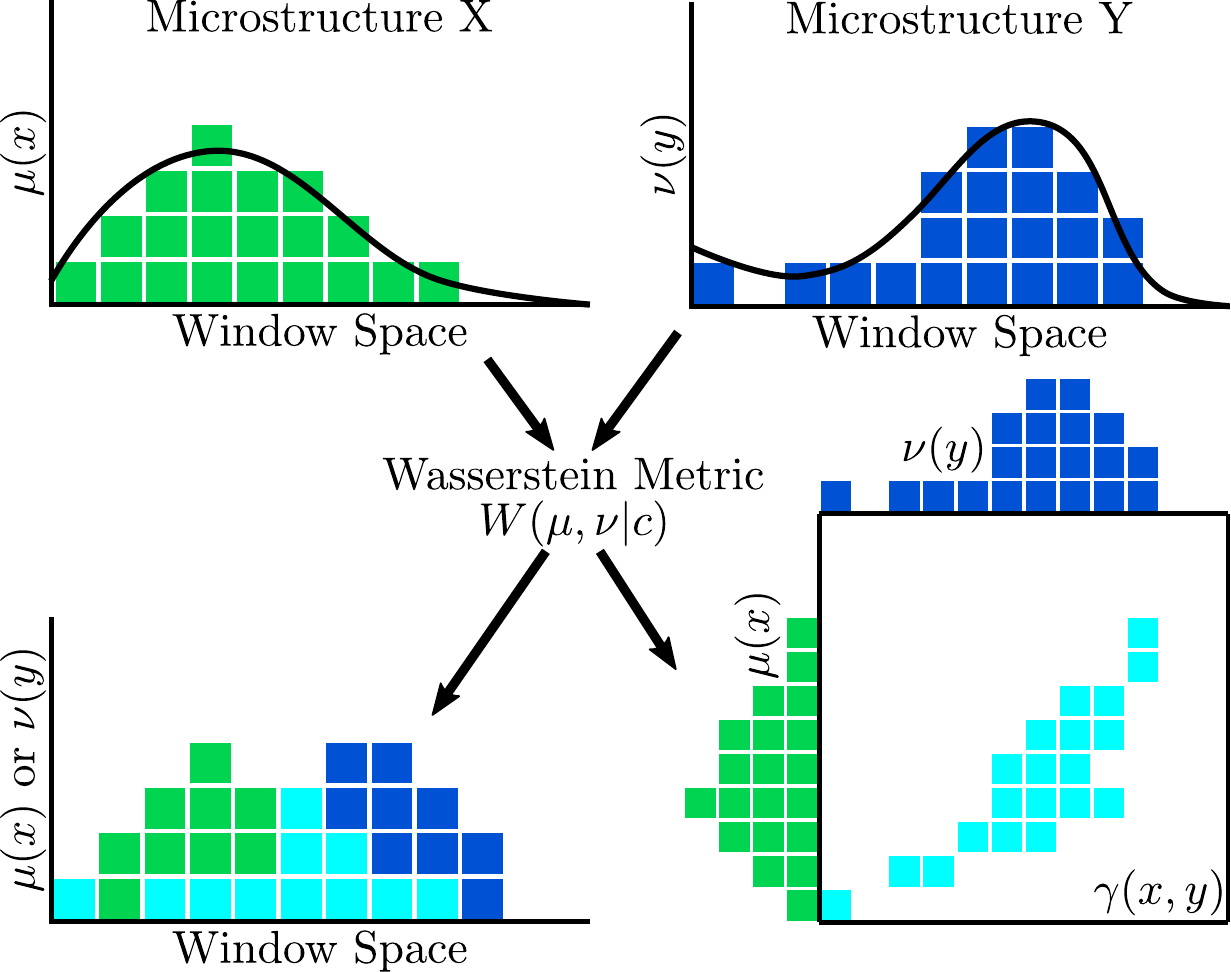}
    \caption{\label{fig:distribution}A schematic showing the use of the Wasserstein metric to compare microstructures $X$ and $Y$.
    The window distributions $\mu(x)$ and $\nu(y)$ (top) are actually two functions on a single window space (bottom left).
    The coupling $\gamma(x, y)$ can be viewed as a transport plan to convert $\mu(x)$ into $\nu(y)$ by transporting the probability mass from a position given by the vertical coordinate to a position given by the horizontal coordinate (bottom right).}
\end{figure}

The situation is represented schematically in Fig.\ \ref{fig:distribution}, with the microstructures of two materials $X$ and $Y$ being represented as probability distributions $\mu(x)$ and $\nu(y)$ on the window space at the top of the figure.
While the underlying probability distributions are continuous they are represented as point clouds of sampled windows, with the histograms indicating the density of points in a given region.
The bottom left panel is intended to emphasize that these are two probability distributions on a single window space, permitting the Wasserstein metric to be used to compare them.
A coupling $\gamma(x, y)$ of $\mu(x)$ and $\nu(y)$ is given in the bottom right panel, with the teal squares indicating that a green square at the vertical coordinate should be coupled to a blue square at the horizontal coordinate.
Viewed as a transportation plan, the contribution to the transportation cost of a single teal square increases linearly with distance from the antidiagonal of the box (a point on the antidiagonal has the same horizontal and vertical coordinates).

While finding an optimal coupling and evaluating the Wasserstein metric can be reduced to a linear programming problem, standard algorithms to solve these have a practical computational complexity of about $O(n^3 \log n)$ where $n$ is the number of points in a point cloud \cite{chvatal1983linear}.
This can be prohibitively expensive when $n$ is in the thousands, limiting the number of windows that can feasibly be sampled and the accuracy of the microstructure representation.
There have recently been efforts in the computer science literature to improve the efficiency of evaluating the Wasserstein metric though, with two approaches of particular relevance here being entropic regularization \cite{cuturi2013sinkhorn,xie2020fastproximal} and mappings to the assignment problem \cite{crouse2016implementing}.

The entropic regularization approach replaces the objective function in Eq.~\ref{eq:optimal} with
\begin{equation}
\mat{\tilde{\gamma}}^* = \argmin_{\substack{\mat{\gamma} \in \Gamma}} \sum_{ij} (c_{ij} + \epsilon \log \gamma_{ij}) \gamma_{ij}
\label{eq:regularized}
\end{equation}
where $\epsilon > 0$ is a regularization parameter.
This problem can be solved using Sinkhorn--Knopp iterations \cite{sinkhorn1967concerning} with a computational complexity of $O(n^2 \log n)$ \cite{altschuler2017near}, giving an efficient way to find an approximate solution to the original problem.
The regularization effectively blurs the resulting coupling, with the accuracy and the computational cost increasing as the regularization parameter is reduced \cite{schmitzer2019stabilized}.
A variety of further modifications including $\epsilon$-scaling \cite{schmitzer2019stabilized} and overrelaxation \cite{thibault2021overrelaxed} have been proposed to reduce the computational cost further.
The authors developed an implementation of the Wasserstein metric that uses many of these modifications to approximate the distance between sets of sampled windows, with the resulting implementation being available in the supplementary materials.

Let $U$ be the set of the elements of $X$ with nonzero weights $\mu(x)$ and $V$ be the set of elements of $Y$ with nonzero weights $\nu(y)$.
Provided that $\mu(x)$ and $\nu(y)$ are such that the optimal transport plan necessarily involves a one-to-one matching of elements of $U$ with elements of $V$, the calculation of the Wasserstein metric can be reduced to an assignment problem \cite{votaw1952methods}.
This can be formulated by constructing a bipartite graph with parts $U$ and $V$ where the weights of the edges are given by the corresponding entries $c_{ij}$ of the cost matrix.
The minimum cost matching (a maximal independent edge set that realizes the minimum total edge weight) is the solution of this assignment problem.
The Jonker-Volgenant-Casta\'non (JVC) algorithm is a standard approach to the assignment problem that finds the solution with a time complexity of $O(n^3)$ where $n$ is the number of nodes of the bipartite graph \cite{crouse2016implementing}.
The JVC algorithm is used here to evaluate the Wasserstein metric when comparing individual windows as described below.

\section{A Metric on Microstructure Windows}
\label{sec:window_comparison}

As discussed in Secs.\ \ref{sec:state_space} and \ref{sec:wasserstein}, a full definition of the metric $W(\mu, \nu | c)$ on the microstructure state space requires a metric $c$ to measure the distance between windows.
There are several reasonable choices of such a metric, but having already introduced the Wasserstein metric in Sec.\ \ref{sec:wasserstein}, we propose to use it for this purpose as well.

\subsection{Grain Boundary Mass Function}
\label{subsec:mass_function}

Since the Wasserstein metric effectively measures the cost of transforming one mass function into another, it is necessary to associate such a mass function with each window.
Let a window be a two-dimensional square grid of pixels with the mass function defined to be one for all pixels adjacent to a grain boundary and zero for all pixels on a grain interior.
This function can be concisely represented as a sparse matrix provided that the average grain size is sufficiently large relative to the pixel spacing to resolve the grain geometries.
Using the notation of Sec.\ \ref{sec:wasserstein}, the window metric is then evaluated by setting $X = \set{x_1,\ldots,x_N}$ and $Y = \set{y_1,\ldots,y_M}$ to be the sets of grain boundary pixels of the first and second windows, respectively, and $\mu_i = 1$ and $\nu_j = 1$ for all $i$ and $j$.
The cost matrix elements are defined to be $c_{i,j} = \abs{x_i - y_j}_{1}$ or the $L^1$ distances between the corresponding pixels.
Assuming that $M = N$, the window distance is the cost of the optimal transport plan given by Eqs.\ \ref{eq:coupling} and \ref{eq:optimal}. 
Concentrating mass along the grain boundaries allows the cost of this transport plan to be effectively viewed as the energy that would be dissipated by moving the grain boundaries from one configuration to the other.
That said, there is no requirement that the optimal transport plan be one where the mass moves contiguously, meaning that strictly interpreting the transport plan along these lines would lead to physical inconsistencies where grain boundaries split or merge.

\begin{figure}
\includegraphics[width=\columnwidth]{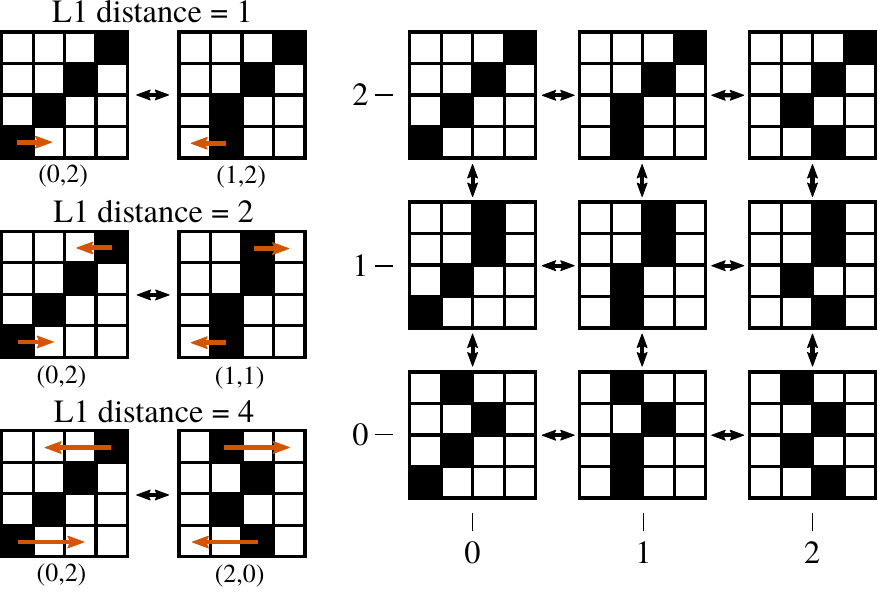}
\caption{\label{fig:window_comparison_figure.png}
Distances induced by the Wasserstein metric using the $L^1$ ground distance between subsets of $4 \times 4$ windows with a total mass of four.
Examples of optimal transport plans (orange vector fields) are shown on the left.
A subset of the window space on the right shows that the Wasserstein metric induces a self-consistent notion of window proximity (i.e., a window space topology).
Each two-sided arrow indicates a distance of one between adjacent windows, and the window at position $(0, 0)$ is a distance of one from $(0, 1)$, two from $(1, 1)$, and four from $(2, 2)$, both in terms of the Wasserstein metric and locations on the grid.}
\end{figure}

Examples are shown in Fig.\ \ref{fig:window_comparison_figure.png} for the comparison of mass functions for $4 \times 4$ windows with a total mass of four.
The Wasserstein metric requires that a distance between points on the underlying space, known as a \emph{ground distance}, be specified.
Here the $L^1$ distance, or the minimum number of steps in cardinal directions required to travel between two pixels, is used for this purpose.
For example, the left side of Fig.\ \ref{fig:window_comparison_figure.png} shows that moving a unit mass one pixel corresponds to a cost of one, moving two unit masses one pixel each corresponds to a cost of two, and moving two unit masses two pixels each corresponds to a cost of four.
Notice that the Wasserstein metric also induces a notion of window proximity; a path between distant windows can be decomposed into a sequence of paths of length one where the mass of a single pixel is moved to an adjacent pixel.
The right side of Fig.\ \ref{fig:window_comparison_figure.png} exhibits this property for a particular set of nine windows, with proximity to one another corresponding both to their distance as computed by the Wasserstein metric and to the $L^1$ distance between the corresponding locations on the grid.
Movement by one step along any of the cardinal directions on the grid requires the movement of a unit mass by one pixel and is accordingly of distance one.
Diagonal neighbors can only be reached by two such steps, resulting in a distance of two.
More generally, the distance between any two windows in the figure is equal to the minimum number of cardinal steps required to travel from one to the other.

More precisely, the right side of Fig.\ \ref{fig:window_comparison_figure.png} reflects that the Wasserstein metric on windows induces a sensible \emph{topology} on the window space, with the meaning that small neighborhoods in the window space around each window overlap in a consistent way.
While it is possible to reduce the window space to only those windows that are distinct up to the symmetries of a square, this would change the definition of the distance, significantly complicate the topology, and would preclude using the window distance to identify anisotropy (e.g., whether grains are elongated along a deposition direction).
It is for these reasons that symmetry operations are not considered here, though if desired there is a version of the Wasserstein metric that can be computed modulo rotations and reflections.
Finally, while the space on the right side of Fig.\ \ref{fig:window_comparison_figure.png} is just a subset of the space of all $4 \times 4$ windows, similar considerations apply to windows containing a larger number of pixels, and distances between windows on the window space of any size continue to align with intuitive notions of window similarity.

\subsection{Unbalanced Wasserstein Metric on Windows}
\label{sec:micrograph_comparison}

As described in Sec.\ \ref{sec:wasserstein}, the conventional Wasserstein metric requires that the mass functions $\mu$ and $\nu$ have the same total mass. 
This constraint is not generally satisfied for the mass functions defined in Sec.\ \ref{subsec:mass_function} since grain boundaries do not obey a conservation law.
This inconsistency is resolved by instead using an unbalanced Wasserstein metric, the difference being that the unbalanced version additionally allows mass to be transported to and matched with a reservoir just beyond the window edge.
The motivation for this is that there could be additional grain boundaries in the micrograph just out of view, meaning that the distance each mass is transported in the optimal transport plan is actually a lower bound for the corresponding transportation distance if the window area were increased.
This construction must be done carefully though to ensure that any unused mass in the reservoir does not appear in the optimal transport plan and affect the value of the unbalanced Wasserstein metric.

\begin{figure}
\centering
\includegraphics[width=\columnwidth]{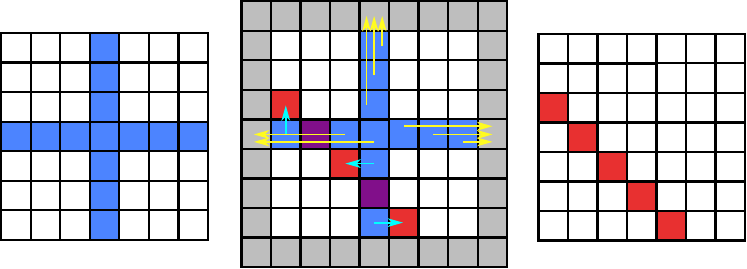}
\caption{\label{fig:nontrivial_transport}
A visual representation of the optimal transport plan between the blue (left) and red (right) grain boundary configurations.
Cyan arrows indicate transport in the window interior, yellow arrows indicate transport to the window boundary, and purple pixels (darkest, center) are not transported.}
\end{figure}

For consistency with the transport of mass within the window, the cost to transport a unit mass to the reservoir is the $L^1$ distance to the closest pixel outside the window;
as shown in Fig.\ \ref{fig:nontrivial_transport}, a mass can either be matched with a mass in the interior of the other window or to the other window's edge, whichever is optimal.
This has the effect of making differences in grain boundary geometry more significant when they occur on the window interiors than close to the window edge since transporting mass already at the window edge to the reservoir is a consistently low cost option.
However, this is not expected to be a significant effect for sufficiently sized windows where the vast majority of mass transport occurs entirely within the window interiors.

By adjusting the mass in the reservoir, the unbalanced Wasserstein metric can formally be written as a modification of the balanced version and solved using classical algorithms.
Specifically, the unbalanced Wasserstein metric is formulated as an assignment problem \cite{crouse2016implementing} that is solved using one of the classical algorithms for this purpose.
This involves constructing a complete bipartite graph with two independent vertex sets corresponding to the unit masses in the two windows, with edge lengths equal to the distances between corresponding pixels.
The vertex set for the first (second) window is augmented by one vertex for each unit mass in the second (first) window, representing the available mass in the first (second) window's reservoir.
Edge lengths between each unit mass in the first (second) window's reservoir and a unit mass on the second (first) window's interior are the shortest $L^1$ distance of the unit mass on the second (first) window's interior to the window boundary.
Finally, edge lengths between masses in the reservoirs of the first and second windows are set to zero to eliminate the effect of any unused reservoir mass. A more formal description of the problem is provided in Appendix \ref{appendix:unbalanced_formal}.

Briefly, a window sampled from a micrograph is binarized with unit masses placed along the grain boundaries.
One window is compared with another by matching grain boundary mass with that of the opposing grain boundary network or by moving the mass beyond the window edge, resulting in a transport plan.
It can be shown that the cost of the optimal transport plan satisfies the properties of a metric, and the resulting quantity is used as a distance between windows.

\section{Microstructure Comparison}
\label{sec:comparing_materials}

\subsection{From Micrographs to Window Distributions}
\label{subsec:micrograph_windows}

Having defined a grain boundary mass function on windows and the unbalanced Wasserstein metric on the window space, the only remaining step to realize the program outlined in Sec.\ \ref{subsec:metric_state_space} is the construction of the window distribution for a given microstructure.
There are three concerns that need to be addressed when doing so.
First, the windows should be sizeable enough to reflect the relevant geometric information about grains and grain clusters.
Second, enough windows should be sampled to adequately characterize the underlying probability distribution on the window space, or at least sufficiently to distinguish distinct microstructures.
Third, the area of the micrograph from which the windows are sampled should be large enough for the sampled windows to be relatively independent.
These three concerns are conflicting though since increasing the area and number of windows sampled from a finite micrograph increases the probability that the windows overlap, introducing statistical dependencies that could bias the apparent distances between microstructures on the microstructure state space.

\begin{figure}[t]
 \includegraphics[width=\columnwidth]{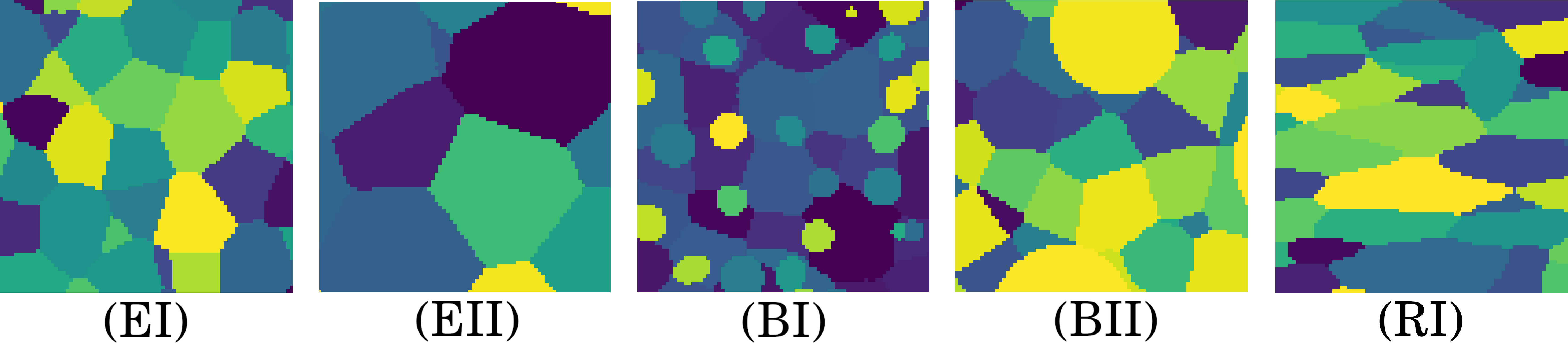}\
\caption{\label{fig:samplepics} Representative microstructures generated by DREAM.3D using the parameters in Table \ref{table:samplestable}.
Micrographs are 80 pixels on a side.} 
\end{figure}

Since the discussion that follows involves empirical tests, five representative microstructures were generated using the DREAM.3D parameters provided in Table \ref{table:samplestable} with the results shown in Fig.\ \ref{fig:samplepics}.
Microstructures EI and EII are equiaxed with average grain diameters of 10 and 20 pixels, respectively.
Microstructures BI and BII are bimodal, with the grains of the matrix resembling those of EI and the grains of the precipitate phase being small and large, respectively.
The grains of microstructure RI have the same volume distribution as those of EI, but have a highly anisotropic morphology with grain diameters in a ratio of $4 \times 2 \times 1$ along orthogonal directions.

\begin{table}[t]
\centering
\caption{\label{table:samplestable} DREAM.3D parameters for the representative microstructures shown in Fig.\ \ref{fig:samplepics}.
All precipitates are equiaxed. ODF stands for the axis ODF weight, E stands for equiaxed, B for bimodal, and R for rolled.}
\newcolumntype{s}{>{\footnotesize}c}\
\begin{tabularx}{\columnwidth} {>{\raggedright}s s s s s s s s}
\hline
Sample &
\multicolumn{5}{s}{Primary Phase} &
\multicolumn{2}{s}{Precipitate}
\\
\hline
&
$\mu$ & $\sigma$ &
B/A & C/A & ODF &
$\mu$ & $\sigma$ 
\\
\hline
EI & 1 & 0.1 & 1.0 & 1.0 & & &  \\
EII & 1.68 & 0.1 & 1.0 & 1.0 & & &   \\
BI & 1 & 0.1 & 1.0 & 1.0 & & 0.3 & 0.1  \\
BII & 1 & 0.1 & 1.0 & 1.0 & & 1.68 & 0.1  \\
RI & 1 & 0.1 & 0.5 & 0.25 & 40000 & &  \\
\hline
\end{tabularx}
\end{table}

\begin{figure*}
    \centering
    \includegraphics[width=\textwidth]{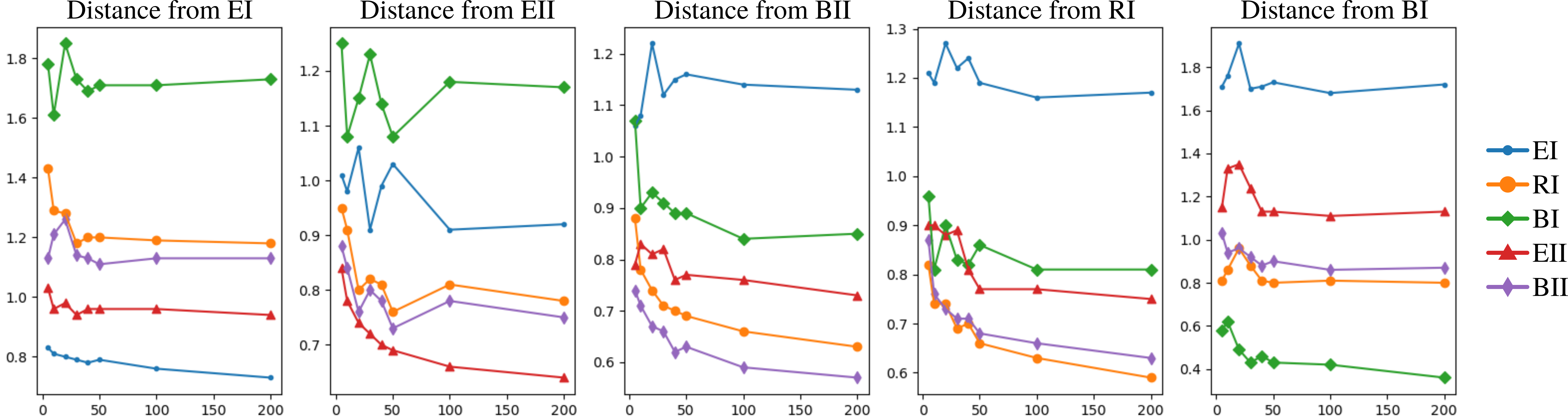}
    \caption{The Wasserstein distance between window distributions (using $40 \times 40$ pixel windows) for the microstructures EI, EII, BII, RI, and BI as a function of the number of sampled windows.
    Each subplot shows the distance from the indicated microstructure to all five microstructures, including to an empirical window distribution independently sampled from the same microstructure.
    The distance of each microstructure to itself is a decreasing function of the number of sampled windows, and presumably converges to zero in the limit.
    Moreover, each microstructure is recognized as being most similar to itself even for small numbers of sampled windows.}
    \label{fig:sampling_stats}
\end{figure*}

Starting with the window size, it is clear that the window area should be a small multiple of the average grain area.
A window that is substantially smaller than a grain would still capture information about the grain boundary inclinations and the grain boundary junction geometries, but would be insensitive to the grain boundary network topology.
Conversely, a window that is substantially larger than a grain would capture considerable geometric information about the grain boundary network, but would result in a probability distribution on the window space that would be impractical to characterize.
Consider that as many as $2^{n \times n}$ distinct mass functions (as defined in Sec.\ \ref{subsec:mass_function}) could exist on a window of $n \times n$ pixels.
The window distribution assigns a probability to each of these mass functions, meaning that a reasonable empirical distribution could require a number of samples that is exponential in the number of window pixels.
While the vast majority of possible mass functions correspond to unphysical grain boundary configurations and realistic window distributions are likely supported on a small fraction of the space, it is nevertheless possible that the number of windows that need to be sampled is still exponential in the window size.
Practically, this means that the window edge length should be two to three times the average grain diameter depending on the grain anisotropy and the morphology of any secondary phases.

With regard to the number of sampled windows, it is useful to begin with an idealized reference case.
Consider an arbitrarily large number of metallographic specimens prepared from the steel plates produced by the facility in Sec.\ \ref{sec:state_space}.
Since all of these steel places were manufactured by a standardized procedure, the windows sampled from each metallographic specimen should be drawn from the same characteristic window distribution.
This means that the corresponding empirical distribution could be constructed to arbitrary accuracy by sampling windows from a sufficiently large area of metallographic specimens.
Moreover, since the underlying window distribution is characteristic of the standardized procedure, the limiting empirical distribution constructed from metallographic specimens provided by a competitor using the same standardized procedure should be identical.
That is, the Wasserstein distance (as defined in Sec.\ \ref{sec:wasserstein}) between the empirical distributions constructed using metallographic specimens produced by the two facilities should converge to zero as the number of specimens increases.

This is unrealistic though because the experimental effort involved limits the number of available metallographic specimens to only a few for any given material.
The limited number of windows that can be sampled causes the Wasserstein distance between the two empirical distributions described above, or even between two empirical distributions independently constructed from metallographic specimens from the same facility, to be nonzero in practice.
This raises the question of the magnitude of the error introduced by limited window sampling, and whether this error affects the reliability of distinguishing different microstructures.
Figure \ref{fig:sampling_stats} directly addresses this question by plotting the Wasserstein distance between pairs of the representative microstructures in Fig.\ \ref{fig:samplepics} as a function of the number of sampled windows.
It is significant that while the distance of each microstructure to itself is consistently nonzero, the self-distance for all microstructures continuously decreases with the number of sampled windows (within stochastic effects).
Moreover, the ordering of the microstructures by similarity is already stable with as few as $100$ sampled windows, and this is a reasonable number to sample experimentally.
The recommendation is therefore to sample a few hundred windows when constructing the empirical window distributions, with the recommended number likely to increase as automated sample preparation and characterization becomes more widely available.

The third consideration is the area of the micrographs from which the windows are sampled.
Consider the limiting case where the area of the micrograph is the same as the area of a single window.
Every window sampled from such a micrograph would necessarily be identical, and increasing the number of sampled windows would not bring the empirical window distribution any closer to the underlying window distribution characteristic of the material.
More generally, an insufficient micrograph area causes the sampled windows to be correlated instead of independently sampling from the underlying window distribution.
Given the previously established effective lower bounds on the window edge length and the number of sampled windows, this sets a practical lower bound on the micrograph area from which windows should be randomly sampled.
An alternative is to reject sampled windows that overlap with previously selected ones, continuing the sampling process until a specified number of disjoint windows is found.
This is the procedure used here, but empirical tests indicate that for the algorithm to be successful the micrograph area should be at least twice the total disjoint window area (e.g., up to $78$ windows of $40 \times 40$ pixels could be sampled from a micrograph of $500 \times 500$ pixels).

\subsection{Comparison of Sample Microstructures}
\label{subsec:comparison_sample}

\begin{figure}[t]
\includegraphics[width=0.75\columnwidth]{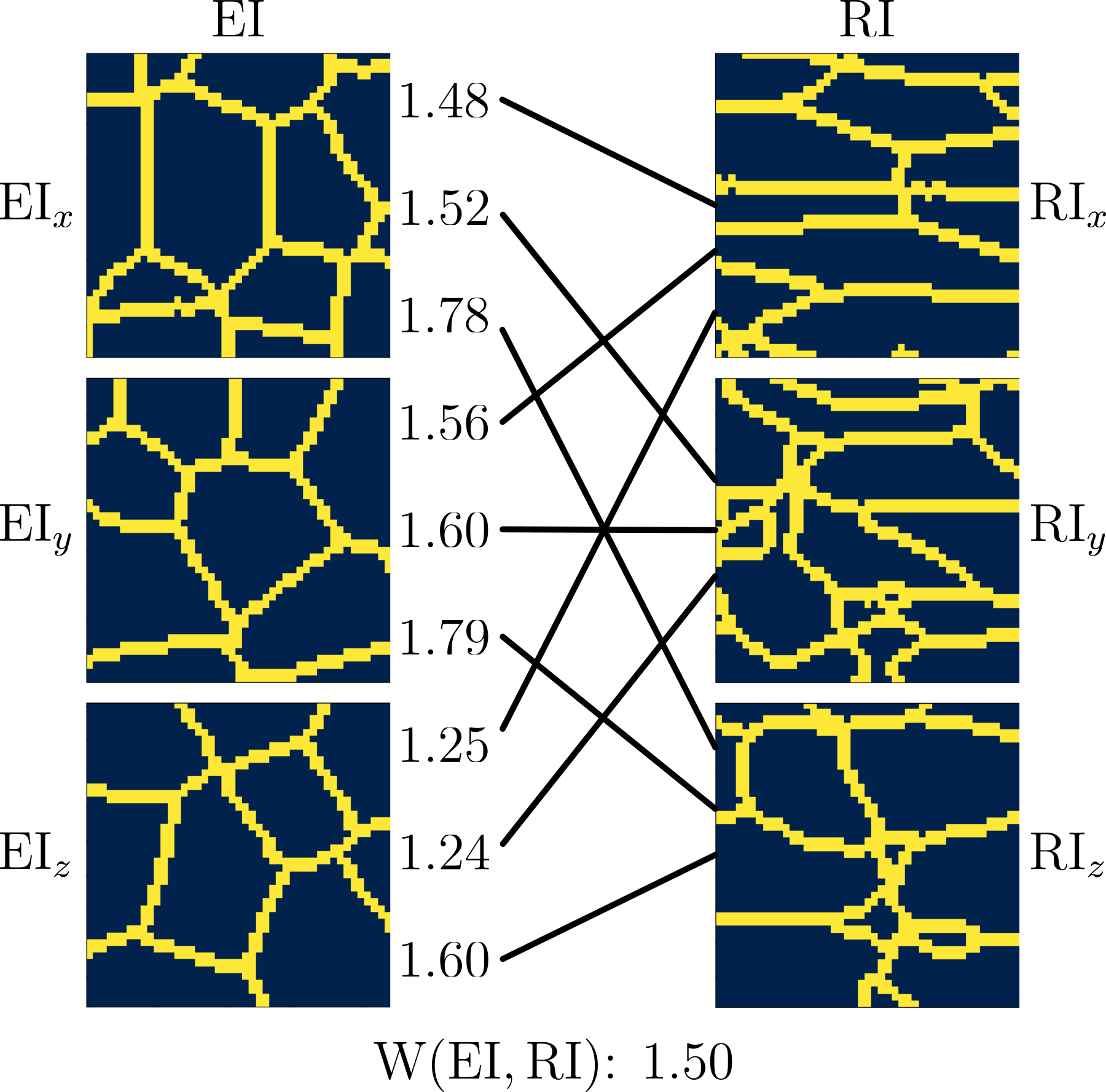}
	\centering
	\caption{\label{fig:E_RIxyz} Empirical window distributions constructed by sampling a single window from each of three orthogonal slices of EI and RI are compared.
	Pairwise distances between windows are shown along the lines connecting the windows and are normalized by the number of window pixels.
	The the optimal pairing of windows (EI$_x$ to RI$_x$, EI$_y$ to RI$_z$, and EI$_z$ to RI$_y$) gives a Wasserstein distance of 1.50.}
\end{figure}

\begin{figure}[t]
    \centering
    \includegraphics[width=\columnwidth]{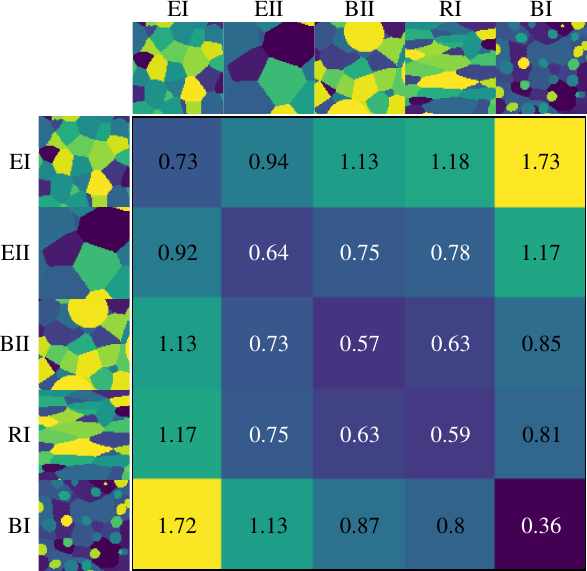}
    \caption{\label{fig:confusion_geometric}Distance matrix for microstructures EI, EII, BII, RI, and BI as generated by DREAM.3D using the parameters in Table \ref{table:samplestable}.
    Empirical window distributions were generated by sampling 200 windows of $40 \times 40$ pixels at random positions by the procedure described in the text.}
\end{figure}

Having established the recommended the window edge length, number of sampled windows, and micrograph area, the approach is now used to compare the microstructures in Fig.\ \ref{fig:samplepics}.
The microstructures generated by DREAM.3D occupy a cubic volume with an edge length of 512 voxels, and sets of two-dimensional slices were sampled at intervals of 10 voxels along each of the three orthogonal axis.
The motivation for collecting slices in this way was to make the resulting empirical window distribution sensitive to anisotropy in the microstructure, and to enable the comparison of anisotropic microstructures while making minimal assumptions about sample alignment.
For example, consider the model empirical window distributions for EI and RI shown in Fig.\ \ref{fig:E_RIxyz}.
These only contain three windows, one sampled perpendicular to each of the coordinate axes.
The pairwise window distances are found as in Sec.\ \ref{sec:window_comparison} and are shown in the figure, and the optimal pairing of windows (EI$_x$ to RI$_x$, EI$_y$ to RI$_z$, and EI$_z$ to RI$_y$) results in an empirical window distribution distance of 1.50.
Aggregating slices in this way provides a combined micrograph area of 6,553,600 pixels from which $200$ windows of $40 \times 40$ pixels are sampled at random positions, ensuring that the sampled windows are reasonably independent.
The Wasserstein distances between the resulting empirical window distributions are shown in the form of a distance matrix in Fig.\ \ref{fig:confusion_geometric}, with the row and column ordering adjusted to group similar microstructures.

Importantly, the diagonal entries are the lowest in each row and column, indicating that the metric correctly identifies each microstructure as being most similar to itself.
However, the diagonal entries are not close to zero and in some situations are only slightly smaller than the off-diagonal entries;
this makes the matrix as constructed not strictly a distance matrix since the identity of indiscernibles is not satisfied.
This occurs because the empirical window distributions are constructed independently for each row and column, meaning that the diagonal entries reflect the magnitude of the error introduced by the limited window sampling.
That said, as long as the self-distance of a microstructure is smaller than the distance to all other microstructures (the off-diagonal entries), then the microstructure is distinguishable from the others.
Notice that the diagonal entry is invariably the smallest entry in the respective row or column, suggesting that the approach can still be used for microstructure identification and classification despite the limited window sampling.
Moreover, the matrix is nearly symmetric, indicating that stochastic errors are likely not substantial for the off-diagonal entries and that the apparent distance between microstructures should be reproducible.
This is reinforced by the overall stability of the distances between dissimilar microstructures for more than $100$ sampled windows in Fig.\ \ref{fig:sampling_stats}.

Observe that EII is the nearest off-diagonal entry to EI, indicating that increasing the grain size and decreasing the grain boundary content is not necessarily as significant as, e.g., the differences in the grain geometry with RI.
BII and RI have different grain size distributions and grain geometries but are at a similar distance to EI, reinforcing that there are many inequivalent ways by which microstructures could be dissimilar.
BI is substantially further from EI than all of the other representative microstructures, likely due to the the smaller effective grain size and higher grain boundary content.
The self-distance of BI is notably smaller than that of any of the other microstructures, and this is interpreted as being due to the relatively shorter distances that grain boundaries need to be transported to transform one window into another in a small-grained microstructure.
Finally, it is noteworthy that the distances separating EII, BII and RI are all comparatively small.
The similarity of EII and BII is reasonable, given that both microstructures contain equiaxed grains and the occasional large precipitates in BII bring the average grain size closer to that of EII.
The similarity of these two with RI is less straightforward though.

\section{A Materials Database}
\label{sec:database}

In principle, the procedures described in Sec.\ \ref{sec:comparing_materials} could be used to construct a polycrystalline materials database with the ability to search for materials that resemble a query presented as an empirical window distribution.
This would not only make existing microstructure data more findable and accessible, but could support the construction of processing routes to, and predicting the properties of, a target microstructure by providing the corresponding data for known materials with similar microstructures.
One obstacle to such a database that has not yet been discussed is the computational cost of a query.
If an empirical window distribution provided as a query contains $n$ sampled windows, the evaluation of the Wasserstein distance defined in Sec.\ \ref{sec:wasserstein} to a single database entry would require the evaluation of $O(n^2)$ pairwise window distances.
For a window containing $m$ pixels, a general-purpose algorithm to calculate a single pairwise window distance as defined in Sec.\ \ref{subsec:mass_function} has an $O(m^3)$ running time.
The result is that for microstructures stored as empirical window distributions with $200$ sampled windows of $40 \times 40$ pixels, a single microstructure comparison would require around five hours on a single processor.
This would make querying a polycrystalline materials database containing even a few hundred materials impractical.

Fortunately, the computational cost can be significantly reduced by changing the way the microstructure comparison is performed.
This relies on the idea that the empirical window distributions in a database entry and a query do not need to contain the same number of sampled windows.
The database entries should certainly contain as many sampled windows as is practical since the purpose of the database is to provide comprehensive microstructure information.
The query could involve fewer windows though, perhaps as few as five, since the random sampling procedure would make these characteristic of a spatially uniform microstructure with high probability.
This effectively converts the question from the similarity of the window distributions of two microstructures into one of the likelihood of the query windows being sampled from the database entry.

\begin{figure*}[ht]
    \centering
    \includegraphics[width=\textwidth]{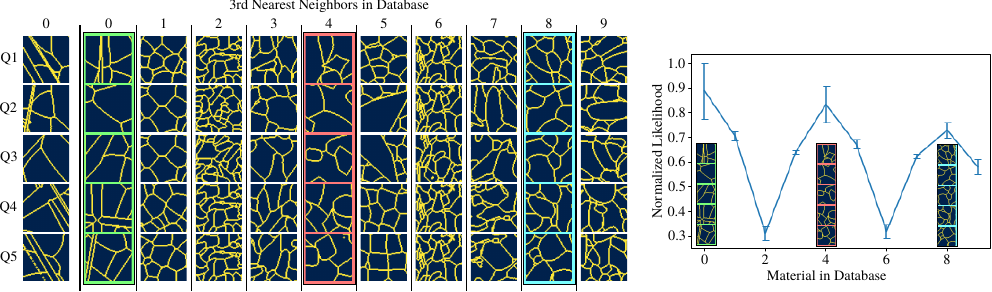}
    \caption{A proof-of-concept database is constructed with microstructures labeled $0$ to $9$ where each entry is represented by $100$ windows of $60 \times 60$ pixels per window.
    The set of five query windows labeled Q1 to Q5 was independently sampled from microstructure $0$.
    The left diagram shows the third-nearest neighbor to each of the query windows for each of the database entries.
    The right diagram shows the inverse distance to the third-nearest neighbor, averaged over the five query windows and scaled by an arbitrary constant.
    Error bars show the standard error of the mean over the five query windows.}
    \label{fig:knn_db}
\end{figure*}

The difficulty with this approach is that the likelihood function is not known; only the empirical window distribution is available.
Nevertheless, the distance of a query window to each of the windows in an empirical window distribution can be evaluated, and the distance of the query window to the $k$th-nearest neighbor in the empirical window distribution is related to the desired likelihood.
Consider that a query window in a low (high) probability density region of the underlying window distribution is likely to have few (many) sampled windows nearby, resulting in a large (small) distance to the $k$th-nearest neighbor.
The use of the $k$th-nearest neighbor instead of the nearest neighbor makes the resulting distance robust to the possible accidental clustering of points in low probability density regions, and is standard practice in distance-based outlier detection models \cite{knox1998algorithms,ramaswamy2000efficient}.
The $k$th-nearest-neighbor distance is inverted to increase (decrease) the score for query windows in high (low) probability density regions, and averaged over the query windows to provide a overall score for the microstructures that most resemble the query.

This approach is used to query a small proof-of-concept synthetic materials database. 
DREAM.3D was used to generate ten materials labeled $0$ to $9$ with randomized morphological statistics (grain size distribution, number of grain types, neighbor number frequency distribution, and shape anisotropy along each axis).
A cubic volume with an edge length of $256$ voxels was generated for each material, and two-dimensional slices were sampled at intervals of $20$ voxels along each of the three orthogonal axis.
The empirical window distribution for each material was generated by sampling $100$ windows of $60 \times 60$ pixels at random locations on randomly-selected two-dimensional slices;
as described in Sec.\ \ref{subsec:micrograph_windows}, the random sampling of windows helps to ensure statistical independence. 
While the number of database entries is quite small, the purpose of this exercise is to show that a database query is possible with a computational complexity that is manageable in practice.

The query was performed by independently sampling a set of five windows labelled Q1 to Q5 from material $0$;
these appear as the leftmost column of Fig.\ \ref{fig:knn_db}.
The distance from each query window to each window of each database entry was calculated, and the third-nearest neighbor to each query window for each database entry is shown in the corresponding row on the left of Fig.\ \ref{fig:knn_db}.
Note that a single window can be the third-nearest neighbor of multiple query windows, e.g., the same window for material $6$ is the third-nearest neighbor for Q1, Q2, Q4 and Q5.

The inverse distance to the third-nearest neighbor, averaged over the five query windows and scaled by an arbitrary constant, is shown on the right of Fig.\ \ref{fig:knn_db};
error bars show the standard error of the mean over the five query windows.
This indicates that the query windows could reasonably have been drawn from database entries 0, 4, or 8, ordered by decreasing likelihood.
This result is intuitively reasonable despite the considerable variability and complexity of the microstructures in the database.
Consider that Q1 and the corresponding third-nearest neighbor for material 0 contain a vertically-oriented lamella among larger grains, and that Q3 and the third-nearest neighbor for material 0 have diagonally-oriented grains on the left and larger grains elsewhere.
Materials 4 and 8 have bimodal grain size distributions, but without the subpopulation of lamella in material 0.
By contrast, materials 1, 3, and 5 have narrower grain size distributions and are identified as being less compatible with the query windows.

While this approach correctly identified the set of five query windows in Fig.\ \ref{fig:knn_db} as being sampled from material 0, it is reasonable to ask whether the computational cost could be reduced further by using fewer query windows.
Preliminary tests indicated that the most likely material was correctly identified only $\about 60\%$ of the time for a single query window, with this increasing to $\about 100\%$ of the time for four query windows.
Five query windows is therefore recommended pending the construction of a more comprehensive materials database for further testing.


\section{Conclusion}
\label{sec:conclusion}

Two principles that arguably underlie integrated computational materials engineering (ICME) are that ongoing advances in processing science enable increased control of material microstructures, and that material properties are predictable functions of material microstructure and chemistry;
there is already evidence that such micro\-struc\-ture-aware material design approaches can converge faster and to better solutions than approaches that relate processing conditions directly to material properties \cite{molkeri2022importance}.
This suggests that any systematic development of ICME capabilities requires a canonical description of material microstructures that is detailed enough to enable accurate prediction of material properties and is general enough to accommodate materials between and beyond existing material classes.
Conventional microstructure descriptors developed for specific material classes (e.g., fiber geometry and volume fraction in fiber-reinforced composites, or size distribution and phase fraction of ferritic and austenitic domains in steels) do not extend across material classes, and machine learning approaches continue to suffer from the comparative scarcity of micrograph data in materials science.

It is proposed that a microstructure can be described as a probability distribution of small volume elements or \emph{windows} at a user-specified length scale, and that this probability distribution can be empirically sampled from micrographs of experimentally-accessible areas.
Provided that there is an available distance function or \emph{metric} that allows the similarity of windows to be quantified, the similarity of two microstructures can be calculated as the cost of transforming one set of sampled windows into the other using the Wasserstein metric.
In an ideal setting, this would allow microstructures to be represented as points in a \emph{microstructure state space} with a well-defined notion of distance, with processing procedures represented as trajectories and material properties represented as scalar or tensor functions on the space.
This concept of a microstructure state space could serve as the foundation on which to build a systematic and general approach to ICME.

The purpose of this paper is to verify that such an approach is indeed possible for the restricted case of single-phase polycrystalline materials (or for general polycrystalline materials where only grain boundary geometry information is retained).
One metric that allows the similarity of windows to be quantified is defined, and practical guidelines for the size of sampled windows, the number of sampled windows, and the area of the micrograph from which to sample are established.
Pairwise distances are evaluated for a set of synthetic microstructures generated by DREAM.3D, and a proof-of-concept microstructure database is constructed with a query capability whose computational cost is manageable.
These efforts are not intended to be authoritative, but merely to show that the vision above could be realized in practice.

Eventually, the metric for the comparison of windows would need to be extended to include phase and crystallographic orientation information for this approach to microstructure representation to be useful as a means of analyzing and guiding experimental efforts.
This is an area of ongoing investigation, but would immediately enable systematic optimization of material microstructures within the context of ICME.

\appendix
\gdef\thesection{\Alph{section}}
\makeatletter
\renewcommand\@seccntformat[1]{\appendixname\ \csname the#1\endcsname.\hspace{0.5em}}
\makeatother

\section{Unbalanced Wasserstein Metric}
\label{appendix:unbalanced_formal}

Let $M$ be a space with a metric $d$ and a distinguished subset $D$ called the boundary.
Given two finite subsets $A,B\subset M\setminus D,$ let $I(A,B)$ be the set of bijections between subsets of $A$ and subsets of $B$. That is,
$$I(A,B)=\set{i: A_0 \to B \mid A_0 \subset A, \text{$i$ is injective}}.$$
The $p$-cost of a matching $i \in I(A,B)$ is
$$C_p(i) = \sum_{\mathclap{a \in A_0}} d[a, i(a)]^p + \sum_{\mathclap{a \in A \setminus A_0}} d(a,D)^p + \sum_{\mathclap{b \in B \setminus i(A_0)}} d(b,D)^p$$
where $d(a, D)$ is the shortest distance from $a$ to any point in $D$.
Then the unbalanced $p$-Wasserstein metric between $A$ and $B$ is defined as
\begin{equation}
\label{eq:wass}
W_p(A,B) = \min_{i \in I(A,B)} [C_p(i)]^{1/p}.
\end{equation}
This is related to Sec.\ \ref{sec:micrograph_comparison} by identifying $M$ with a square grid of pixels, $D$ with the boundary pixels, $A$ and $B$ with mass functions (interpreted as indicator functions) on the windows being compared, and $I(A, B)$ with the set of viable transport plans.

$W_p$ can be computed by constructing an equivalent balanced optimization problem.
Suppose that $A$ contains $m$ points and $B$ contains $n$ points.
Let $\mat{X}$ be the $(m+n) \times (n+m)$ matrix
\begin{equation*}
X_{ij} = 
\left\{
\begin{alignedat}{3}
&d(a_i,b_j)^p \quad& 1 &\leq i \leq m & 1 &\leq j \leq n\\
&d(a_{i},D)^p & 1 &\leq i \leq m & n &< j \leq n+m \\
&d(b_{j},D)^p & m &< i \leq m + n \quad& 1 &\leq j \leq n \\
&0 & m &< i \leq m + n & n &< j \leq m+n
\end{alignedat}
\right.
.\end{equation*}
That is, $\mat{X}$ is a matrix whose first $m$ rows correspond to points of $A$ and first $n$ columns correspond to points of $B$.
Columns $n + 1$ through $n + m$ correspond to dummy points that allow points in $A$ to be matched to the boundary rather than to a point in $B$.
Rows $m + 1$ through $m + n$ serve a similar purpose for points in $B$.
The distance between dummy points is set to zero to avoid any additional cost for matching points in $A$ to points in $B$.
A solution to the assignment problem with cost matrix $\mat{X}$ provides the optimal matching in Eq.\ \ref{eq:wass}.
As such, $W_p(A,B)$ can be found by applying the Jonker-Volgenant-Casta\'non (JVC) algorithm (or equivalent) to the matrix $\mat{X}$.

\section*{Acknowledgements}
B.S.\ is grateful for the support of the National Science Foundation under Award No.\ 2232967.
D.M.\ and J.K.M.\ are grateful for the support of the National Science Foundation under Award No.\ 2232968.

\bibliographystyle{elsarticle-num}
\bibliography{refs}

\end{document}